\documentclass[twocolumn]{webofc}

\usepackage[varg]{txfonts}   
\usepackage{color}
\usepackage{listings}
\usepackage{setspace}
\definecolor{Code}{rgb}{0,0,0}
\definecolor{Decorators}{rgb}{0.5,0.5,0.5}
\definecolor{Numbers}{rgb}{0.5,0,0}
\definecolor{MatchingBrackets}{rgb}{0.25,0.5,0.5}
\definecolor{Keywords}{rgb}{0,0,1}
\definecolor{self}{rgb}{0,0,0}
\definecolor{Strings}{rgb}{0,0.63,0}
\definecolor{Comments}{rgb}{0,0.63,1}
\definecolor{Backquotes}{rgb}{0,0,0}
\definecolor{Classname}{rgb}{0,0,0}
\definecolor{FunctionName}{rgb}{0,0,0}
\definecolor{Operators}{rgb}{0,0,0}
\definecolor{Background}{rgb}{0.98,0.98,0.98}
\lstdefinelanguage{Python}{
numbers=left,
numberstyle=\footnotesize,
numbersep=1em,
xleftmargin=1em,
framextopmargin=2em,
framexbottommargin=2em,
showspaces=false,
showtabs=false,
showstringspaces=false,
frame=l,
tabsize=4,
basicstyle=\ttfamily\small\setstretch{1},
backgroundcolor=\color{Background},
commentstyle=\color{Comments}\slshape,
stringstyle=\color{Strings},
morecomment=[s][\color{Strings}]{"""}{"""},
morecomment=[s][\color{Strings}]{'''}{'''},
morekeywords={import,from,class,def,for,while,if,is,in,elif,else,not,and,or,print,break,continue,return,True,False,None,access,as,,del,except,exec,finally,global,import,lambda,pass,print,raise,try,assert},
keywordstyle={\color{Keywords}\bfseries},
morekeywords={[2]@invariant,pylab,numpy,np,scipy},
keywordstyle={[2]\color{Decorators}\slshape},
emph={self},
emphstyle={\color{self}\slshape},
}

\usepackage{hyperref}
\usepackage{cleveref}
\begin{document}
\title{A computational EXFOR database}
%
%

\author{\firstname{Georg} \lastname{Schnabel}\inst{1}\fnsep\thanks{Currently at IAEA NAPC-Nuclear Data Section} \thanks{\email{g.schnabel@iaea.org}}}

\institute{
Division of Applied Nuclear Physics, Uppsala University, Sweden
}

\abstract{%
  The EXFOR library is a useful resource for many people in the field of nuclear physics.
  In particular, the experimental data in the EXFOR library serves as a starting point for nuclear data evaluations.
  There is an ongoing discussion about how to make evaluations more transparent and reproducible.
  One important ingredient may be convenient programmatic access to the data in the EXFOR library from high-level languages.
  To this end, the complete EXFOR library can be converted to a MongoDB database.
  This database can be conveniently searched and accessed from a wide variety of programming languages, such as C++, Python, Java, Matlab, and R.
  This contribution provides some details about the successful conversion of the EXFOR library to a MongoDB database and shows simple usage examples to underline its merits.
  All codes required for the conversion have been made available online and are  open-source.
  In addition, a Dockerfile has been created to facilitate the installation process.
}
\maketitle
\section{Introduction}
\label{intro}
The EXFOR library~\cite{otukaMoreCompleteAccurate2014} as a comprehensive collection of experimental reaction data is a valuable resource for many people in the field of nuclear physics.
The Nuclear Reaction Data Centers (NRDC) host online services to search for data, visualize, and retrieve them in various formats, e.g., among those the well known EXFOR web retrieval system~\cite{zerkinExperimentalNuclearReaction2018}.
In particular, both the flexible search options provided by those online services and the simple structure of so-called computational formats, such as C4, are tremendously helpful in nuclear data evaluations.
Notwithstanding the indisputable value of existing services and formats, there may be use cases involving reaction data from EXFOR which are not optimally covered by existing services and formats yet:

\begin{enumerate}
\item
There is an ongoing discussion in the community about how to make evaluations more transparent and reproducible, as demonstrated by the recent proposal for WPEC subgroup 49 on reproducibility in nuclear data evaluation,
and the idea of automated evaluation pipelines gains momentum.
Such a pipeline is a sequence of scripts in a programming language to perform an evaluation.
Choices of an evaluator are implemented as instructions in scripts, thereby removing the need for manual intervention in the event that an evaluation needs to be reproduced.
The creation of such scripts should be facilitated as much as possible to enable evaluators to focus on the essential evaluation work.
Also readability counts: Other evaluators should later be able with as little effort as possible to understand the scripts.
Convenient access in just a few lines of code to the EXFOR library, which means both searching and retrieving data, from a variety of programming languages serves this goal.

\item
It can be foreseen that more and more sophisticated methods from statistics and machine learning will be used to analyze the data in the EXFOR library.
The variety of available algorithms and the rapid development of new ones suggest that it may be difficult to come up with a computational format that suits all potential requirements.
In this application scenario, the user should be put in the position to quickly create a customized computational format themselves and traverse the EXFOR library in any way they want. 
\end{enumerate}

The solution put forward in this contribution to enable user-friendly access to information in the EXFOR library is to convert the complete EXFOR library to a MongoDB database~\cite{MongoDBDatabaseModern} using JSON~\cite{JSONFormat}  as an intermediate format.
This database application and the associated database architecture are very popular in the commercial world.
It can be downloaded and used free of charge.
Due to their widespread adoption, MongoDB databases can be accessed from a wide variety of programming languages including C++, Python, Java, Perl, Matlab and R.
Being endowed with a rich query language and the possibility to access any textual and non-textual information suggest that a MongoDB database is a viable solution for the two described use case scenarios above.

Finally, we want to stress that similar efforts have been undertaken in the past and associated codes released to the public to facilitate accessing data in the EXFOR library:
The X4TOC4 code~\cite{X4toC4Cullen} developed by D.E. Cullen converts EXFOR entries into a tabular format.
The ENDVER code~\cite{ENDVERTrkov} developed by A. Trkov can convert EXFOR data to the computational C4 format. 
The \textit{x4i} code developed by D.A. Brown provides a programmatic interface to EXFOR, and a fork of it, the \text{x4i3} code~\cite{x4i3} is available on GitHub.
A thorough comparison to these codes and their functionality and design considerations is outside the scope of this paper.
Also the web retrieval system of the IAEA has been upgraded recently to enable the retrieval of data in the JSON format.

In this contribution, we first discuss the original EXFOR format~\cite{EXFORBasicsReport,EXFORSystemsManual,EXFORBasicsShort} in comparison with the JSON format~\cite{JSONFormat}.
Afterwards we discuss the creation of a MongoDB database based on EXFOR data given in the JSON format.
Then we provide a simple example of how the MongoDB EXFOR database can be accessed from Python.
A conclusion section ends this document.

\section{From EXFOR to JSON}
\label{sec:from_EXFOR_to_JSON}

The EXFOR format~\cite{EXFORBasicsReport,EXFORSystemsManual,EXFORBasicsShort} is a data exchange format designed to enable the storage of numerical data and textual information related to experiments.
Data associated with an experiment is bundled together as a so-called \textit{entry}.
An entry comprises several \textit{subentries}.
The first subentry contains general bibliographical information, such as the authors of the measurements or the facility, and specifications common to all subsequent subentries.
From the second subentry onwards, each subentry contains the data associated with a specific reaction process measured in the experiment.

The EXFOR format defining the structure of an entry was conceived to be readable by humans and machines.
In order to be readable by machines, it follows rigid structural rules.
For instance, a line contains two or more fields and the size of a field must be a multiple of eleven characters.
Maximal 66 character slots in a line can be used to accommodate fields.
Many lines are dedicated to store key-value pairs:
The first field of size eleven specifies a keyword, such as \verb|AUTHOR|, and the second field of size 55 contains the associated value, e.g., the names of the authors.

The rigid structure is in principle a good thing as it facilitates to write programs to parse the content in EXFOR entries, making the content accessible in a high-level programming language.
However, the syntax of the EXFOR format follows complex rules and reflects its coevolution with the Fortran programming language.
Writing parsers for EXFOR entries in a high-level language, such as Python or R, is therefore a tedious and time-consuming endeavor. 

Furthermore, even though the EXFOR format is human-readable, modifications by hand bear the risk of inadvertently violating format specifications.
For instance, the \verb|BIB| keyword introducing a block with bibliographical information and the description of the data is followed by two fields indicating the number of keywords and the number of lines in the block.
The entry is easily put into an inconsistent state by adding a bibliographical field by hand without changing accordingly such \textit{counter} fields.
Here one may argue that users do not need to change EXFOR entries themselves.
However, it can be countered that format specifications may also be motivated by the needs of users.
If users have the option to create their individual EXFOR entries with customized fields, e.g., an alternative representation of covariance matrices,  pertinent in their domain of application, this could provide helpful input for potential extensions of the official EXFOR library in the future.

Both problems, the need to write a parser and less error-prone modifications by hand, can be solved by converting the entries given in the EXFOR format to another format that is widely supported in the field of information technology.
A natural candidate for that purpose is the JSON format~\cite{JSONFormat}.
It is a hierarchical format and can store numerical and textual data.
Therefore it provides all the features to store without any loss of information or accuracy all the information available in an EXFOR entry.

An R package to convert entries or subentries given in the EXFOR format to the JSON format has been implemented and is available for download~\cite{ExforParserEXFORJSON}.
The philosophy of the converter is to preserve the logical structure of the original EXFOR entry as much as possible.
Keywords in the original entry are also keywords in the JSON object.

Once an EXFOR entry or subentry is available as JSON object, one can make use of existing functionality in most of the popular programming languages to retrieve information or manipulate the JSON object.
Here is a simple example that shows how to extract information from an EXFOR entry given as JSON object using Python:
\begin{lstlisting}[language=Python]
import json
with open('exforEntryFile') as json_file:
    entry = json.load(json_file)
    
entry['SUBENT'][0]['BIB']['AUTHOR']
entry['SUBENT'][3]['BIB']['REACTION']
entry['SUBENT'][3]['DATA']['TABLE']
\end{lstlisting}

As this example shows, having the EXFOR entries stored as JSON objects greatly facilitates the access to the information.
Also the creation of customized computational formats becomes doable by users without too much efforts, e.g., by writing a small Python script to this end.
However, it is equally important to be able to effectively search for relevant data, which is the topic of the next section.

\section{About the MongoDB database}
\label{sec:about_MongoDB}

Storing a large collection of data and searching through them is the purpose of database software.
Often so-called \textit{relational databases} are employed for storing data.
The data is organized in several tables and the association of rows residing in different tables is established by columns containing the same information even though in different tables.
For instance, in the context of EXFOR, each table could possess a column \verb|EntryID| containing the entry identification numbers which uniquely identify experiments.
This column links the rows of different tables together.

Using this database type to store the EXFOR library requires to restructure the information present in the collection of EXFOR entries as a collection of tables.
The variability in terms of the amount of information of different subentries makes this conversion process challenging.
For instance, a specific field type can be present in one subentry but absent in another one.
These challenges for the conversion can and have been solved, e.g., as has been proven by the conception and implementation of the web retrieval system and its underlying SQL database of the IAEA~\cite{zerkinExperimentalNuclearReaction2018}.

As the organization of data in EXFOR entries (hierarchical) is different from that in an SQL database (relational), there may be use cases where it is beneficial to preserve the original organization of the data if the benefit of powerful search capabilities associated with a database application can be achieved as well. 
Database solutions that can store the EXFOR entries in a way that preserves their original structure are available in the form of so-called \textit{document-oriented databases}.
This database type belongs to the class of \textit{NoSQL} databases.
Instead of using a collection of tables, document-oriented databases manage a collection of documents.
A document includes all information related to one entity.
This is in sharp contrast to relational databases where properties belonging to one entity are usually distributed over several tables.
The advantage of this database software in context of the EXFOR library is that each EXFOR entry can be stored as a document without the need to restructure or distribute its information.

The database adopted here is MongoDB~\cite{MongoDBDatabaseModern}, which can be downloaded and used free of charge.
In a MongoDB database, documents are organized in collections.
Documents are stored in the BSON format which stands for \textit{binary JSON}.
Technical details aside, the BSON and JSON format are equivalent.
Due to this reason, it is possible to directly insert EXFOR entries given in the JSON format into a collection of a MongoDB database.
A script to convert EXFOR entries to JSON objects and insert them into a MongoDB database has been made available at~\cite{ExforParserEXFORJSON}.

Once the MongoDB database is filled, the expressive query language provided by the MongoDB database software can be used to search for relevant information.
Simple usage examples showing how EXFOR data can be queried from Python will be the topic of the next section.
The remainder of this section elaborates on some choices made for the conversion from the original EXFOR library to a MongoDB database as it is provided at~\cite{DockerfileInstallationInstructions}.
The following list describes effected modifications, which are in the opinion of the author reasonable and helpful:

\begin{enumerate}

\item
The logical unit most people operate with are not entries but subentries.
For this reason, the documents in the MongoDB database are subentries.

\item 
The first subentry of an entry contains information which is common to all subsequent subentries.
For this reason, the information of the first subentry has been merged into all subsequent subentries of the same entry, before they were added to the MongoDB database.
Sometimes collisions of field names occur during the merging process.
In such cases, the offending field names in the first subentry are altered by adding the suffix \verb|_firstSub| to them prior to merging.
\item
The information in a \verb|COMMON| section of an original EXFOR subentry contains quantities that are constant for all measured data points.
For instance, an angle differential cross sections may have been measured at various angles for 15\,MeV incident neutrons.
Thus the neutron energy is often stored in a \verb|COMMON| field to avoid redundancy.
From the user point of view, it may be still helpful to have the incident energy stored in the \verb|DATA| table together with the angles, and the measured cross section value.
Information in the \verb|COMMON| section is therefore merged into the \verb|DATA| table but nevertheless also preserved as a separate field.

\item
Standardized units are helpful to remove conversion errors.
Therefore all energies have been converted to MeV and all cross sections to millibarn.
Also units of compound quantities such as associated with angle differential cross sections and spectra are modified accordingly.

\end{enumerate}

Some other modifications of less significance have not been mentioned here for the sake of brevity.
They are documented in the manual at~\cite{DockerfileInstallationInstructions} accompanying the installation files for the MongoDB EXFOR database.

\section{Simple usage example}
\label{sec:usage_examples}
According to the TIOBE index~\cite{TIOBEIndex} one of the most popular programming languages is Python, which finds also broad adoption in the field of nuclear physics.
Therefore some simple examples how to retrieve data from the MongoDB EXFOR database are provided here to demonstrate the ease of use.
The following examples rely on the \verb|pymongo| module.

To interact with the database, one needs first to connect to it:
\begin{lstlisting}[language=Python]
from pymongo import MongoClient
client = MongoClient('localhost', 27017)
db = client["exfor"]
entries = db["entries"]
\end{lstlisting}

One of the most elementary user actions is to retrieve a subentry using its subentry identification number (an eight digit string).
\begin{lstlisting}[language=Python, firstnumber=5]
subent = entries.find_one({'ID':'11701004'})
\end{lstlisting}
The expression passed as an argument to the function \verb|find_one| specifies the search query.
Search queries for a MongoDB database have to be formulated as JSON objects.
Since the data structure called a (nested) \textit{dictionary} in Python is essentially equivalent to a JSON object, pythonists can probably get used to this query syntax quickly.

The result of the query is a (potentially nested) dictionary.
It can be explored by making use of the Python functions provided for dictionaries.
Just as an example:
\begin{lstlisting}[language=Python, firstnumber=6]
subent["BIB"]["AUTHOR"]
subent["DATA"]["UNIT"]
subent["DATA"]["TABLE"]
\end{lstlisting}

As a final example for a more advanced query, we can use a regular expression to match reaction strings that specify neutrons (\verb|N|) as projectile, Fe-56 as target, and angle-integrated cross sections (\verb|SIG|):
\begin{lstlisting}[language=Python, firstnumber=9]
import re
regex = re.compile(
  "^\(26-FE-56\(N,[^)]+\)[^,]*,,SIG\)"
)
subents = entries.find(
  { 'BIB.REACTION' : regex }
)
\end{lstlisting}

The variable \verb|subents| is an iterator, which can be used to iterate over the found subentries in a loop, e.g.:
\begin{lstlisting}[language=Python, firstnumber=16]
for subent in subents:
  print(subent["BIB"]["AUTHOR"])
\end{lstlisting}

Finally, the connection to the MongoDB database should be closed:
\begin{lstlisting}[language=Python, firstnumber=18]
client.close()
\end{lstlisting}

This example provided just a small glimpse into the possibilities to interact with the data in the MongoDB EXFOR database.
Comprehensive information about the query language can be found in the official MongoDB documentation.

\section{Conclusions}
\label{sec:conclusion}

We argued that the JSON format is a suitable format to store all the information available in the EXFOR library.
Entries and subentries in the EXFOR library can be converted without loss of information or accuracy to the JSON format.
Due to the wide support for the JSON format, the extraction of EXFOR data from JSON objects is trivial in high-level languages, such as Python.
An EXFOR to JSON converter package has been made available at~\cite{ExforParserEXFORJSON}.

We also argued that a relational database may not always be the ideal solution to store the data of the EXFOR library from the perspective of the user.
Document-oriented databases, such as MongoDB, enable storing the EXFOR library without structural transformations.
As JSON objects can be imported into a MongoDB database and an EXFOR to JSON converter is in place, the conversion of the complete EXFOR library into a MongoDB database is straight-forward.
A script that automates this conversion process is available at~\cite{CreateExforDbScriptCreate}.

The complete installation process of the MongoDB EXFOR database requires several steps, such as the installation of the MongoDB database and the conversion of EXFOR entries.
Therefore, to facilitate the installation for the user, the installation process has been automated to a large extent using the Docker technology for virtualization.
The required files and installation instructions to install the EXFOR MongoDB on the local computer can be found at~\cite{DockerfileInstallationInstructions}.

It is the hope that the provided computational database will be helpful for users in its current form.
Future use cases will certainly point to possible improvements and issues will potentially surface.
Because modifications of the database can be effected by the user themselves without too much efforts, users can become designers and for instance create their own computational formats and databases.
This circumstance may foster the development of tools and formats related to nuclear data and potentially provides inspiration for how to improve the official EXFOR library. 
\vspace{2ex}

%
\bibliography{paper}

\begin{thebibliography}{14}

\bibitem{otukaMoreCompleteAccurate2014}
N.~Otuka, E.~Dupont, V.~Semkova, B.~Pritychenko, A.~Blokhin, M.~Aikawa,
  S.~Babykina, M.~Bossant, G.~Chen, S.~Dunaeva et~al., Nuclear Data Sheets
  \textbf{120}, 272 (2014)

\bibitem{zerkinExperimentalNuclearReaction2018}
V.~Zerkin, B.~Pritychenko, Nuclear Instruments and Methods in Physics Research
  Section A: Accelerators, Spectrometers, Detectors and Associated Equipment
  \textbf{888}, 31 (2018)

\bibitem{MongoDBDatabaseModern}
\emph{{{MongoDB}}: {{The}} database for modern applications},
  \url{https://www.mongodb.com/}

\bibitem{X4toC4Cullen}
D.~Cullen, A.~Trkov, Tech. Rep. IAEA-NDS-0080, IAEA (2001)

\bibitem{ENDVERTrkov}
A.~Trkov, Tech. Rep. IAEA-NDS-77, IAEA (2008)

\bibitem{x4i3}
\emph{{x4i3} - {EXFOR} interface for {P}ython},
  \url{https://github.com/afedynitch/x4i3}

\bibitem{EXFORBasicsReport}
Tech. Rep. IAEA-NDS-0206, IAEA (2008)

\bibitem{EXFORSystemsManual}
V.~McLane, Tech. Rep. IAEA-NDS-207, IAEA (2000)

\bibitem{EXFORBasicsShort}
\emph{{{EXFOR Basics}}: {{A}} short guide to the neutron reaction data exchange
  format}, \url{https://www-nds.iaea.org/nrdc/basics/}

\bibitem{JSONFormat}
\emph{{{JSON}} format}, \url{http://www.json.org/}

\bibitem{ExforParserEXFORJSON}
\emph{{{exforParser}}: {{EXFOR}} to {{JSON}} converter},
  \url{https://github.com/gschnabel/exforParser}

\bibitem{DockerfileInstallationInstructions}
\emph{Dockerfile and installation instructions to create {{MongoDB EXFOR}}
  database}, \url{http://www.nucleardata.com/\#dockerimages}

\bibitem{TIOBEIndex}
\emph{{{TIOBE Index}}}, \url{https://www.tiobe.com/tiobe-index/}

\bibitem{CreateExforDbScriptCreate}
\emph{{{createExforDb}}: {{Script}} to create {{MongoDB}} database with
  {{EXFOR}} data}, \url{https://github.com/gschnabel/createExforDb/}

\end{thebibliography}
%
%
%
%

\end{document}